\documentclass[twocolumn,showpacs,prl,superscriptaddress,amsmath,floatfix]{revtex4}
 
\usepackage{graphicx}

\newcommand{\figurewidth}{\columnwidth}

\begin{document}

\title{Strong universality and algebraic scaling in two-dimensional Ising spin glasses}

\author{T.~J\"{o}rg} \affiliation{Dipartimento di Fisica, Universit\`a
  di Roma ``La Sapienza'', SMC and INFM, P.le.~Aldo Moro 2, 00185
  Roma, Italy.}

\author{J.~Lukic} \affiliation{Dipartimento di Fisica, Universit\`a di
  Roma ``La Sapienza'', SMC and INFM, P.le.~Aldo Moro 2, 00185 Roma,
  Italy.}

\author{E.~Marinari} \affiliation{Dipartimento di Fisica, INFN,
  Universit\`a di Roma {\em La Sapienza}, P.le Aldo Moro 2, 00185
  Roma, Italy.}

\author{O.~C.~Martin} \affiliation{Laboratoire de Physique Th\'eorique
  et Mod\`eles Statistiques, b\^atiment 100, Universit\'e Paris-Sud,
  F--91405 Orsay, France.}
 
\date{\today}

\begin{abstract}
At zero temperature, two-dimensional Ising spin glasses are known to
fall into several universality classes. Here we consider the scaling
at low but non-zero temperature and provide numerical evidence that
$\eta \approx 0$ and $\nu \approx 3.5$ in all cases, suggesting a
unique universality class.  This algebraic (as opposed to exponential)
scaling holds in particular for the $\pm J$ model,
with or without dilutions and for the plaquette diluted model.
Such a picture, associated with an exceptional
behavior at $T=0$, is consistent with a real space renormalization
group approach. We also explain how the scaling of the specific heat
is compatible with the hyperscaling prediction.
\end{abstract}


\maketitle


\paragraph*{Introduction ---}
Critical phenomena in disordered systems remain a severe challenge. In
fact, many properties that are well established in pure systems (such
as upper critical dimension, universality, etc.) are still open to
debate when strong disorder is present. This is the case in particular
for spin glasses (SG), i.e., magnetic systems incorporating both
disorder and frustration~\cite{MezardParisi87b}.

In this work we focus on two-dimensional SG because (i) they
are computationally tractable and (ii) their zero temperature ($T=0$)
universality classes~\cite{AmorusoMarinari03} are well established.
For our purposes there are two such $T=0$ classes, associated with
whether or not all excitation energies are multiple of a given
quantum. The corresponding spin glass stiffness exponent $\theta$ is
zero in the first class (as in the $\pm J$ model) and is close to
$-0.28$ in the second class (as when the couplings are Gaussian random
variables).  Following standard scaling
arguments~\cite{FisherHuse86,BrayMoore87} to infer the finite
$T$  behavior, the thermal exponent should be given by
$\nu=-1/\theta$; when $\theta=0$, formally $\nu = \infty$ so that the
correlation length $\xi$ should diverge faster than any power of
inverse temperature $\beta=1/T$ (e.g., exponential scaling).  One's
expectation is then that the two universality classes defined at $T=0$
could have direct counterparts at $T>0$.

Recent numerical work~\cite{HartmannHoudayer04,KatzgraberLee04} has
confirmed the relation $\nu=-1/\theta$ in the
case of Gaussian couplings; similarly, all of the latest
works~\cite{SwendsenWang87,Houdayer01,LukicGalluccio04,KatzgraberLee05}
on the $\pm J$ model (for which $\theta=0$) have concluded that $\nu =
\infty$. We reconsider here this issue in greater depth, comparing
models with different distributions of spin-spin couplings, using
advanced Monte Carlo and partition function solvers, and applying
sophisticated extrapolations to the infinite volume limit. In
contrast with previous claims, we conclude that $\xi(T)$ diverges with
the same exponent $\nu \approx 3.5$ for all considered models!
Furthermore, we find that the exponent $\eta$ also seems to be the
same regardless of the details of the couplings.  We are thus in a
situation where the $T>0$ properties fall into \emph{just one
universality class}, whereas two classes arise when considering the
$T=0$ properties (where the quantized nature of energies can be
relevant).  A simple interpretation of this unusual phenomenon is that
at $T=0$ one has an ``additional'' fixed point (realized only when
energies are quantized) but that this fixed point is irrelevant for
the critical properties defined just above $T_c$.  
We have also investigated the specific heat singularities
of our models. Here we find a behavior that is asymptotically
compatible with $T_c=0$ hyperscaling.

\paragraph*{Models, computational methods and analysis ---}

We consider the $2D$ Edwards-Anderson model defined by the
Hamiltonian~\cite{EdwardsAnderson75}
\begin{equation}
  \label{eq:H_EA}
  {\cal H} = - \sum_{\langle xy \rangle} J_{xy} \sigma_x \sigma_y\; ,
\end{equation}
where the sum is over all the nearest neighbor pairs of a $2D$ square
lattice of length $L$ with periodic boundary conditions. The 
$\sigma_x = \pm 1$ are Ising spins. We take the couplings 
$J_{xy}$ to be independent random variables, chosen from
several distributions in order to check their effect on the
critical behavior. In particular, we shall consider:
(1) the $\pm J$ model where $J_{xy}=\pm 1$ with equal
probability; (2) the diluted $\pm J$ model obtained from 
the undiluted case by setting a fraction $(1-p)$ of the couplings
to zero; (3) the ``irrational model'' where 
$J_{xy}=\pm 1$ or $\pm G$ with equal probability for these
4 values, $G$ being the golden mean, 
$G=(1 - \sqrt{5})/2 \approx -0.618$; (4) A distribution with $\frac12$ of
the couplings equal to $\pm 1$ and one half equal to $\pm\frac14$: we
call it ``Gap 1/4''.

\begin{figure}[htbf]
  \includegraphics[width=0.7\figurewidth]{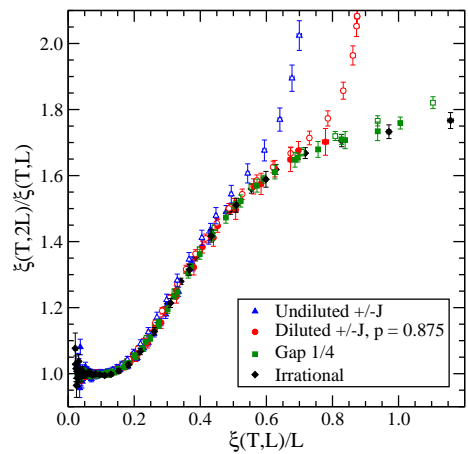}
  \caption{\label{fig:step_scaling} The FSS function $F_\xi$.
}
\end{figure}

We tackled these systems using two complementary approaches.  First,
we applied cluster replica Monte Carlo along with parallel tempering
and numerous algorithmic
optimizations~\cite{SwendsenWang86,HukushimaNemoto96,Houdayer01}; such
an approach can provide high quality estimates of configurational
averages even for large lattices. Second, we used exact
partition function computations~\cite{GalluccioLoebl00} which give us
the value of the free energy and specific heat for individual samples
at arbitrary temperatures.  In all cases, for each lattice size $L$,
we used a large number of disorder realizations for each of the
distributions mentioned previously; typically, we used thousands of
such samples.
In spite of the large lattices sizes we are able to handle,
it is appropriate to use sophisticated techniques for analyzing
the effects of finite $L$. Starting with 
a suitably defined finite-volume correlation length $\xi(T,L)$
(see later)
and a long-range observable ${\cal{O}}(T,L)$ such 
as the spin glass susceptibility $\chi_{SG}(T,L)$,
finite size scaling (FSS) theory predicts
\begin{equation}
  \label{eq:FSS_1}
  \frac{{\cal O}(T,s L)}{{\cal O}(T,L)}   =
  F_{{\cal O}} \Bigl( \xi(T,L)/L; s \Bigr) + O\Bigl(\xi^{-\omega},L^{-\omega}\Bigr)\;.
\end{equation}
Here, $F_{{\cal O}}$ is the FSS function and $s>1$ is a scale factor. 
Eq.~(\ref{eq:FSS_1}) is an excellent starting
point for investigations of the FSS behavior, as it involves
only finite-volume quantities taken from a pair of systems with sizes
$L$ and $s L$ at a given $T$. The knowledge of the scaling
functions $F_{{\cal O}}$ (where $\cal O$ is in our case $\chi_{SG}$)
and $F_{{\xi}}$ (where $\cal O$ is $\xi$) allows us to extract
information on the critical behavior using an infinite volume
extrapolation \cite{CaraccioloEdwards95,PalassiniCaracciolo99}.  This
technique works with data strictly above $T_c$ and hence is well
suited to our case for which $T_c=0$. In essence it uses $F_{{\cal
    \xi}}$ and $F_{{\cal O}}$ to obtain the thermodynamic limit of
${\cal O}$ using an iterative procedure in which the pair $\xi$ and
${\cal O}$ is scaled up from $L \to s L \to s^2 L \to \ldots \to
\infty$ as described in \cite{CaraccioloEdwards95}. Detailed
knowledge of the critical behavior can then be obtained from
appropriate fits to the extrapolated data.  To show that this
relatively sophisticated technique works well in our system, we
display in Fig.~\ref{fig:step_scaling} the FSS function $F_{\xi}$
determined from these procedures, for several of our models.  The
excellent data collapse at small values of $\xi/L$ ($\xi/L < 0.45$)
(already noticed in \cite{CheungMcMillan83a,CheungMcMillan83b})
validates the FSS framework.  Furthermore we see that these FSS
functions are independent of the distribution of couplings $J_{xy}$;
that is precisely what should transpire if there is a single
universality class in our system.  The region of the
excellent data collapse increases with the size of the system
suggesting strongly a single limiting FSS function. 
The diluted and the Gap 1/4 model are the most important pieces of
evidence on which our conclusions rest.  The diluted model has smaller
finite size effects than the undiluted one: this allows us to go to very
reliable extrapolations. We have been able to thermalize systems
up to $L=64$ (with $2000$ samples).

\paragraph*{The spin-glass susceptibility and the exponent $\eta$ ---}

In our Monte Carlo simulations we follow two replicas from which 
we can measure the local spin overlaps 
$q_{xy} = \sigma_{xy}^{(1)} \sigma_{xy}^{(2)}$
(the superscript is the replica index). Of interest is
the two site correlation function of these overlaps; we define in fact
the wave-vector-dependent spin-glass susceptibility via
\begin{equation}
  \label{eq:susceptibility}
  \chi_{SG}({\vec k}) = \frac{1}{L^2} \sum_{{\vec r_1}}\sum_{{\vec r_2}} 
e^{i {\vec k} \cdot ({\vec r_2} - {\vec r_1})} 
[\langle q_{\vec r_1} q_{\vec r_2} \rangle_{\rm T} ]_{\rm av}\;.
\end{equation}
The usual spin-glass
susceptibility is then defined through 
$\chi_{SG}(T,L) = L^2 \langle q^2 \rangle = \chi_{SG}({\vec k}={\vec 0})$.
(We denote the thermal average at
temperature $T$ by $\langle \ldots \rangle_{\rm T}$ and the average over
the disorder realizations by $[ \ldots ]_{\rm av}$.)

\begin{figure}[htbf]
  \includegraphics[width=0.65\figurewidth]{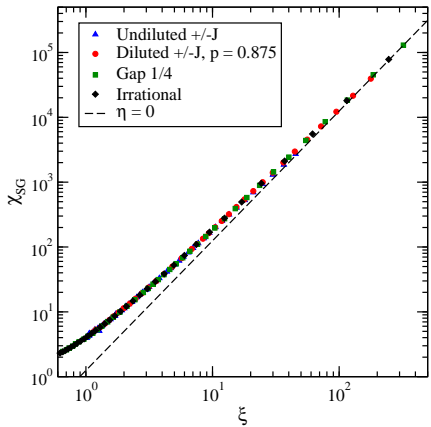}
  \caption{\label{fig:chi_vs_xi_2D} Parameter-free plot 
of $\chi_{SG}(T,\infty)$ versus $\xi(T,\infty)$ for 
different coupling distributions. The line is for $\eta=0$.}
\end{figure}

The spin-glass susceptibility is a measure of the correlation 
volume of our system; using the standard form for a
correlation function near a critical point,
\begin{equation}
G(r) = \frac{e^{-r/\xi}}{r^{D-2+\eta}}
\end{equation}
we see that in $D=2$  as $T \to 0$
$\chi_{SG}$ should behave as
$\chi_{SG} \sim \xi^{2-\eta}$. To check this,
we use our infinite volume extrapolations for both 
$\xi(T)$ and $\chi_{SG}(T)$, and display one versus the other
in Fig.~\ref{fig:chi_vs_xi_2D}.
It is remarkable that the data for our models fall on the same curve,
indicating in particular that they all have the same
exponent $\eta$, a highly surprising fact if there were two 
universality classes. Fits of this curve lead to
values of $\eta$ that are very small, between
$0$ and $0.1$, strongly suggestive of $\eta=0$; note that one expects the
models with continuous couplings to have $\eta=0$
exactly.

\paragraph*{The correlation length and the exponent $\nu$ ---}

\begin{figure}[htbf]
  \includegraphics[width=0.7\figurewidth]{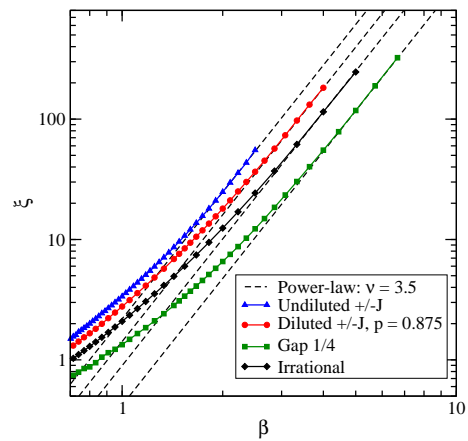}
  \caption{\label{fig:xi_vs_T_2D} $\xi(\beta,\infty)$ as a function of $\beta$.}
\end{figure}

The second moment of the overlap correlation function can be
identified with the square of the correlation length. In practice, we
define this length via 
\begin{equation}
  \label{eq:correlation_length}
  \xi(T,L) = \frac{1}{2 \sin(k_{min}/2)} \left[ \frac{\chi_{SG}(0)}{\chi_{SG}(k_{min})} - 1 \right]^{\frac{1}{2}} \,,
\end{equation}
where $\chi_{SG}$ is defined in Eq.~(\ref{eq:susceptibility}).  Note
that $k_{min}= 2 \pi/L$ is the modulus of the smallest non-zero
wave-vector allowed by periodic boundary conditions.  We use our
analysis methods to extract the large $L$ limit of
$\xi(T,L)$. Having done so for our different models, we find that
their temperature dependence is not so different. This is illustrated
in Fig.~\ref{fig:xi_vs_T_2D}; we also see that the
different models lead to rather parallel asymptotes, suggesting that
the exponent $\nu$ is the same for all of them.  Analogous plots on a
semi-log scale do not give evidence for an exponential scaling of
$\xi$, even for the $\pm J$ model.  Finally, performing power fits, we
find values of $\nu$ close to $3.5$.  Again, this goes in the
direction of a single universality class.

\paragraph*{The specific heat and the exponent $\alpha$ ---}

We measured the specific heat density $c_v$ both via Monte Carlo 
and via the exact partition function techniques.
In contrast to what was observed for $\chi$ and $\xi$, 
the $\pm J$ model and its diluted versions differ 
from the continuous models when one considers 
the very low $T$ behavior of $c_v$. In the first class of models,
$c_v$ decreases fast as one approaches $T=0$, while in the
second class, $c_v$ looks linear in $T$. According to 
hyperscaling, when $T_c=0$, one has the relation\cite{BakerBonner75}
\begin{equation}
c_v(T) \sim T^{- \alpha} \quad {\rm with} \quad \alpha = -2 \nu
\label{hyper}
\end{equation}
We saw that $\nu \approx 3.5$, so we expect $\alpha \approx -7$.
The deceptive 
linear behavior in the continuous model
can be easily understood. In the
ground state of a continuous model, the local field on a spin
is a continuous variable with a finite density at 0. Thus
at any low temperature, a fraction proportional to $T$ of
the system's spins can be thermally activated, thereby leading to 
a linear dependence of $c_v$ in $T$. The point is that the exponent
$\alpha$ describes the non-analytic scaling of
$c_v$, but one can also have regular parts, namely linear, quadratic
etc. in $T$. Since $-\alpha$ is large, the non-analytic part is
sub-dominant compared to the analytic contributions and thus
numerically invisible. The conclusion is that the continuous models
make it impossible in practice to estimate $\alpha$. Fortunately,
the situation is completely different for the $\pm J$ model 
and its variants. There the leading analytic contribution to $c_v$
is of the form $T^{-2}\exp(-4J/T)$ because the local field on any 
spin is a multiple of $2J$. This is far smaller 
than the predicted non-analytic term (a power of $T$), so we
are able to use these models to estimate $\alpha$.
\begin{figure}[htbf]
  \includegraphics[width=0.65\figurewidth,angle=-90]{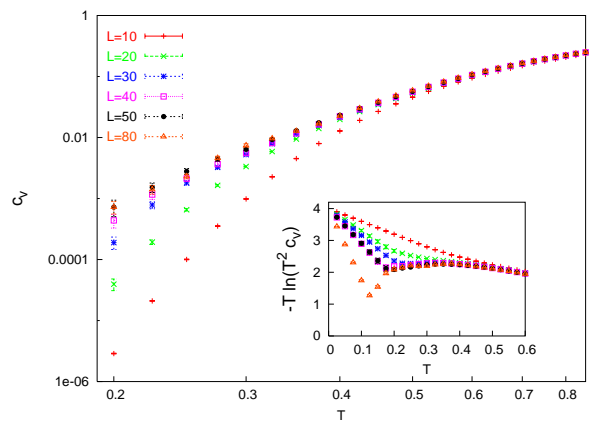}
  \caption{\label{fig:cv_vs_beta_2D} Specific heat of
  the $\pm J$ model versus $T$ in a log-log plot.
  Inset: plot of $-T \ln(T^2 c_v)$ versus $T$.}
\end{figure} 
In Fig. \ref{fig:cv_vs_beta_2D} we show the specific heat of the $\pm
J$ model versus $T$.  In the inset we show the
semi-logarithmic plot of $c_v$ to test for exponential scaling. 
The data are not compatible with an exponential scaling
(the $L=80$ data are crucial to reach this conclusion), while an
asymptotic power law behavior is consistent with the data and with the
hyperscaling of Eq. (\ref{hyper}).  $2D$ disordered systems with
different choices of a (frustrated) disorder behave in the same
way\cite{Lukic06}, making our claim about a strong universality even
stronger.

\paragraph*{A renormalization group justification ---}

In the picture we are proposing
the thermal properties ($T>0$)
of $2D$ SG fall into just one universality
class, in contrast to the properties defined at exactly $T=0$.
Furthermore, the relation $\nu=-1/\theta$ holds \emph{if} we take
$\theta$ from the $T=0$ class 
of ``continuous models'' (where $\theta$ is non-zero).
We are thus faced with a situation where continuous models
behave as expected, but the $\pm J$ models
have (i) a separate (exceptional) behavior at $T=0$;
(ii) fall into the continuous class when $T>0$.
To justify theoretically this scenario, we appeal to a real space
renormalization group (RG) framework. Consider the Migdal-Kadanoff (MK)
construction for spin glasses~\cite{SouthernYoung77}. These
hierarchical lattices can be generated by recursively applying an
``inflation'' operation whereby each edge
is expanded into a cell having $b$ parallel branches
of length $s$.
Renormalization on these lattices can be done exactly.
For the
purpose of this discussion, consider the case where $b=s=3$.
When
the distribution of couplings is continuous, one has $\theta \approx
-0.278$, while when using the $\pm J$ distribution one has
$\theta=0$. Indeed there are \emph{two} fixed point
distributions of the couplings: a generic (continuous) one for which
the energy scale goes to zero ($\theta < 0$) and another one which is
very special, being a sum of delta functions on odd integers (and
leading to $\theta=0$).

\begin{figure}[htbf]
  \includegraphics[width=0.65\figurewidth,angle=270]{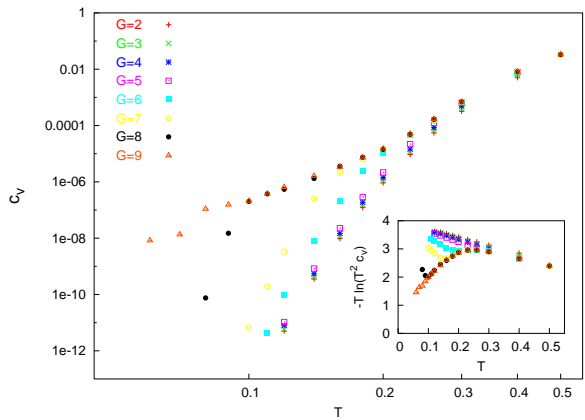}
  \caption{\label{fig:cv_vs_beta_MK_2D} Specific heat for the
  $\pm J$ model on
  MK lattices versus $T$ in a log-log display.
$G$ is the number of recursions of the MK construction.
  Inset: same data but for a semi-log display,
  which should be straight in the case of exponential scaling.} 
\end{figure}

Now what happens when we turn on the temperature in this real space
RG framework?  At very low $T$ one is very close
to the critical manifold so the renormalization will first flow toward
one of the $T=0$ fixed points; the corresponding fixed
point determines the critical exponents of the system.  In the
case of the $\pm J$ model, the initial distribution starts by being
nearly concentrated on odd integers. However, as we increase the
lattice size, this initial distribution will be less and less of that
form: the iterations renormalize the \emph{free} energy, $F =
E-TS$, and since $S$ fluctuates, we loose the quantization property
rapidly with size.  The flow is thus toward the ``continuous''
distribution fixed point for which $\theta<0$.  In
Fig.~\ref{fig:cv_vs_beta_MK_2D} we display the specific heat as a
function of $T$ for this $\pm J$ model; as expected, the
behavior indicates a power scaling of $c_v$ totally compatible with
the value of $\alpha$ obtained from the continuous distribution
fixed point.

\paragraph*{Summary and conclusions ---}

In this study of two dimensional SG, we examined
the effect of the underlying distribution of couplings on
the critical thermodynamics. We found several remarkable behaviors:
(1) $\eta$ is model insensitive and very close to zero
in all models. 
(2) the divergence of $\xi$ does not seem to depend
on the model, and the corresponding exponent 
is given by $\nu = -1/\theta$ as long as one uses
the $\theta$ of the continuous models.
(3) The specific heat exponent $\alpha$ can be measured
using the discrete models, whereas 
in the continuous models this singularity is sub-dominant
and thus numerically inaccessible. Hyperscaling is consistent
with the data.

From our observations we conclude that thermal properties of $2D$ SG
fall into a single universality class. This is in contrast to what
happens exactly at $T=0$: there, an additional fixed point of the
RG arises, allowing different scaling in the
continuous and in the $\pm J$ type models.

\acknowledgments
This work was supported by the EEC's FP6
IST Programme under contract IST-001935, EVERGROW, and by the EEC's
HPP under contracts HPRN-CT-2002-00307 (DYGLAGEMEM) and
HPRN-CT-2002-00319 (STIPCO).  The LPTMS is an Unit\'e de Recherche de
l'Universit\'e Paris~XI associ\'ee au CNRS.

\bibliographystyle{prsty}

\bibliography{references}

\addcontentsline{toc}{chapter}{\protect\bibname}
\begin{thebibliography}{10}

\bibitem{MezardParisi87b}
M. M{\'e}zard, G. Parisi, and M.~A. Virasoro, {\em Spin-Glass Theory and
  Beyond} (World Scientific, Singapore, 1987).

\bibitem{AmorusoMarinari03}
C. Amoruso, E. Marinari, O.~C. Martin, and A. Pagnani, Phys. Rev. Lett. {\bf
  91},  087201  (2003).

\bibitem{FisherHuse86}
D.~S. Fisher and D.~A. Huse, Phys. Rev. Lett. {\bf 56},  1601  (1986).

\bibitem{BrayMoore87}
A.~J. Bray and M.~A. Moore, Phys. Rev. Lett. {\bf 58},  57  (1987).

\bibitem{KatzgraberLee04}
H. Katzgraber, L. Lee, and A. Young, Phys. Rev. B {\bf 70},  014417  (2004).

\bibitem{HartmannHoudayer04}
A. Hartmann and J. Houdayer, Phys. Rev. B {\bf 70},  014418  (2004).

\bibitem{SwendsenWang87}
R.~H. Swendsen and J.-S. Wang, Phys. Rev. Lett. {\bf 58},  86  (1987).

\bibitem{Houdayer01}
J. Houdayer, Eur. Phys. Jour. B {\bf 22},  479  (2001).

\bibitem{LukicGalluccio04}
J. Lukic {\it et~al.}, Phys. Rev. Lett. {\bf 92},  117202  (2004).

\bibitem{KatzgraberLee05}
H. Katzgraber and L. Lee, Phys. Rev. B {\bf 71},  134404  (2005).

\bibitem{EdwardsAnderson75}
S.~F. Edwards and P.~W. Anderson, J. Phys. F: Met. Phys. {\bf 5},  965  (1975).

\bibitem{SwendsenWang86}
R.~H. Swendsen and J.-S. Wang, Phys. Rev. Lett. {\bf 57},  2607  (1986).

\bibitem{HukushimaNemoto96}
K. Hukushima and K. Nemoto, J. Phys. Soc. Jpn. {\bf 65},  1604  (1996),
  cond-mat/9512035.

\bibitem{GalluccioLoebl00}
A. Galluccio, M. Loebl, and J. Vondr\'ak, Phys. Rev. Lett. {\bf 84},  5924
  (2000).

\bibitem{CaraccioloEdwards95}
S. Caracciolo {\it et~al.}, Phys.~Rev.~Lett. {\bf 74},  2969  (1995).

\bibitem{PalassiniCaracciolo99}
M. Palassini and S. Caracciolo, Phys. Rev. Lett. {\bf 82},  5128  (1999).

\bibitem{CheungMcMillan83a}
H.-F. Cheung and W.~L. McMillan, J. Phys. C: Solid State Phys. {\bf 16},  7027
  (1983).

\bibitem{CheungMcMillan83b}
H.-F. Cheung and W.~L. McMillan, J. Phys. C: Solid State Phys. {\bf 16},  7033
  (1983).

\bibitem{BakerBonner75}
G.~A. Baker and J.~C. Bonner, Phys. Rev. B {\bf 12},  3741  (1975).

\bibitem{Lukic06}
J. Lukic, E. Marinari, and O.~C. Martin, Europhys. Lett. {\bf 73},    (2006).

\bibitem{SouthernYoung77}
B.~W. Southern and A. Young, J. Phys. C {\bf 10},  2179  (1977).

\end{thebibliography}

\end{document}